\documentclass[lettersize,journal]{IEEEtran}
\IEEEoverridecommandlockouts

\usepackage{cite}
\usepackage{amsmath,amssymb,amsfonts}
\usepackage{algorithmic}
\usepackage{graphicx}
\usepackage{textcomp}
\usepackage{xcolor}
\usepackage{algorithm}
\usepackage{array}
\usepackage[caption=false,font=normalsize,labelfont=sf,textfont=sf]{subfig}
\usepackage{textcomp}
\usepackage{stfloats}
\usepackage{url}
\usepackage{verbatim}
\usepackage{cite}
\usepackage{amssymb}
\usepackage{bm}
\usepackage{cuted}
\usepackage{subcaption}
\usepackage{placeins}

\usepackage[utf8]{inputenc}
\usepackage{lineno}
\usepackage{cases}
\usepackage{float}
\usepackage{bbold}
\usepackage{siunitx}

\newcommand\bW{\ensuremath{{\bold W}}}

\newcommand\by{\ensuremath{{\bm y}}}
\newcommand\bG{\ensuremath{{\bold G}}}
\newcommand\bh{\ensuremath{{\bm h}}}

\newcommand\bP{\ensuremath{{\bold P}}}

\newcommand\bC{\ensuremath{{\bold C}}}

\newcommand\ba{\ensuremath{{\bm a}}}

\newcommand\bF{\ensuremath{{\bold F}}}

\newcommand\bD{\ensuremath{{\bold D}}}

\newcommand\bPhi{\ensuremath{{\bm \Phi}}}

\newcommand\btheta{\ensuremath{{\bm \theta}}}

\newcommand\bY{\ensuremath{{\bold Y}}}

\newcommand\bU{\ensuremath{{\bold U}}}

\newcommand\bS{\ensuremath{{\bold S}}}

\newcommand{\diag}{\mathrm{diag}}

\newcommand\bQ{\ensuremath{{\bold Q}}}

\newcommand{\bI}{{\bold I}}

\usepackage{amsthm}
\usepackage{xcolor}

\def\BibTeX{{\rm B\kern-.05em{\sc i\kern-.025em b}\kern-.08em
    T\kern-.1667em\lower.7ex\hbox{E}\kern-.125emX}}

\begin{document}
	\bstctlcite{BSTcontrol}
	
	\title{Cram\'er-Rao Bound Minimization for Discrete SIM-aided ISAC Systems}

	\author{Xuejun Cheng,~\IEEEmembership{Graduate Student Member,~IEEE,} Yuhui Jiao, Qian Zhang,~\IEEEmembership{Member,~IEEE,} \\Jiancheng An,~\IEEEmembership{Senior Member,~IEEE,} Maoyuan Wang, Runlai Wang, and~Ju Liu,~\IEEEmembership{Senior Member,~IEEE}
		\thanks{\scriptsize This work was supported in part by the National Natural Science Foundation of China under Grant 62071275 and 62471096.
		Xuejun Cheng and Yuhui Jiao	contributed equally to this work. The corresponding authors: Ju Liu; Zhang Qian.  }
		\thanks{\scriptsize Xuejun Cheng, Yuhui Jiao, Maoyuan Wang, Runlai Wang, and Ju Liu are with the School of Information Science and Engineering, Shandong University, Qingdao, 266237, China (email: \{chengxuejun, yuhuijiao2024, maoyuanwang2024, 202312661\}@mail.sdu.edu.cn; juliu@sdu.edu.cn).}
		\thanks{\scriptsize Qian Zhang is with School of Computer and Communication Engineering, Northeastern University at Qinhuangdao, Qinhuangdao 066004, China (e-mail: zhangqian@neuq.edu.cn).}
		\thanks{\scriptsize Jiancheng An is with the School of Electronic Science and Engineering, University of Electronic Science and Technology of China (UESTC), Chengdu, 611731, China (e-mail: jiancheng.an@uestc.edu.cn).}
	}
	
	
	\maketitle

\begin{abstract}

Stacked intelligent metasurfaces (SIM) have recently emerged as a promising wave-domain signal processing architecture for integrated sensing and communication (ISAC), owing to the capability of efficient beamforming with reduced number of radio-frequency chains. However, most existing SIM-aided ISAC designs rely on maximizing beampattern gain under continuous phase shifts, which cannot explicitly characterize the fundamental estimation accuracy under practical deployment. In this paper, we investigate Cramér–Rao bound (CRB) minimization problem for a discrete SIM-aided ISAC system. 
To tackle the resulting non-convex problem, we develop a two-stage transmit design that decouples CRB optimization from cascaded multi-layer SIM realization. In the first stage, an ideal precoder is obtained. In the second stage, it is synthesized on the practical SIM by alternating power allocation and phase-shift updates, with a closed-form discrete projection and linearized QoS constraints. Numerical results demonstrate fast convergence, reveal the sensing-communication trade-off, and verify CRB improvements with higher phase resolution and larger SIM configurations.

\end{abstract}
\vspace{-5pt}
\vspace{-12pt}
\section{Introduction}
\vspace{-3pt}
Integrated sensing and communication (ISAC) has been recognized as a key enabling paradigm for sixth-generation (6G) wireless networks \cite{Zhang2026RISISAC}. 
By exploiting the shared spectrum, hardware, and waveform resources, ISAC aims to perform information transmission and target sensing in a unified radio system.  However, fully digital arrays require a large number of radio-frequency (RF) chains, leading to increasing hardware cost and power consumption \cite{Cheng2026RobustRISRSMA}. 
To enhance spatial flexibility, various emerging reconfigurable architectures have recently been investigated \cite{11595178,an2024sim_mimo}.
Recently, stacked intelligent metasurfaces (SIM) provide a new wave-domain signal processing architecture with fewer transmit RF chains for addressing the above challenge \cite{an2023sim_hmimo,an2024sim_mimo,zhang2026spherical}. An SIM consists of multiple reconfigurable transmissive metasurfaces, which  offer additional spatial degrees of freedom for beamforming and radiation-pattern control \cite{11534974,an2024sim_mimo}. Motivated by the above observations, several studies have investigated SIM-aided ISAC systems. Niu \textit{et al.} investigated a downlink multiuser SIM-ISAC system under sensing power constraint \cite{niu2024sim_isac}. Li \textit{et al.} considered a transmit SIM-aided ISAC architecture and shaped the normalized sensing beampattern while maximizing the user sum rate \cite{li2025sim_isac_tvt}. Zhang \textit{et al.} studied SIM-aided ISAC system while enhancing the illumination toward the target \cite{zhang2025discrete_sim_isac}. On the other hand, Wang \textit{et al.} introduced Cram\'er-Rao bound (CRB) for target response matrix estimation considering extended-target sensing under continuous phase shift SIM \cite{wang2024multiuser_sim_isac}. These studies either adopt radiation power sensing metrics or assume continuous phase shifts, leaving direct angular CRB minimization under discrete SIM phase constraints insufficiently explored.

In this paper, we investigate a discrete SIM-aided multiuser ISAC system, where a base station equipped with an SIM simultaneously serves multiple communication users and estimates the direction of a point-like target. The formulated problem is highly non-convex due to the fractional structure of the CRB objective, the cascaded coupling of the layers and the discrete phase constraints together with the quality of service (QoS) constraints. To address this problem, we propose a two-stage transmit design framework that decouples the CRB optimization from the cascaded multi-layer SIM realization. 
In the first stage, we derive an optimal precoder via a semidefinite relaxation (SDR) convex reformulation. In the second stage, this precoder is synthesized on the practical SIM by alternately updating the power allocation and phase shifts, where the non-convex QoS constraints are handled via successive convex approximation (SCA). Numerical results demonstrate substantial CRB reduction while satisfying QoS requirements, reveal the sensing-communication trade-off, and show the benefits of larger SIM apertures and higher phase-shift resolution for sensing accuracy.

\vspace{-7pt}
\section{System Model}
\vspace{-4pt}
\label{s2}

\begin{figure}[!t]
	\centering
	\includegraphics[width=0.45\textwidth]{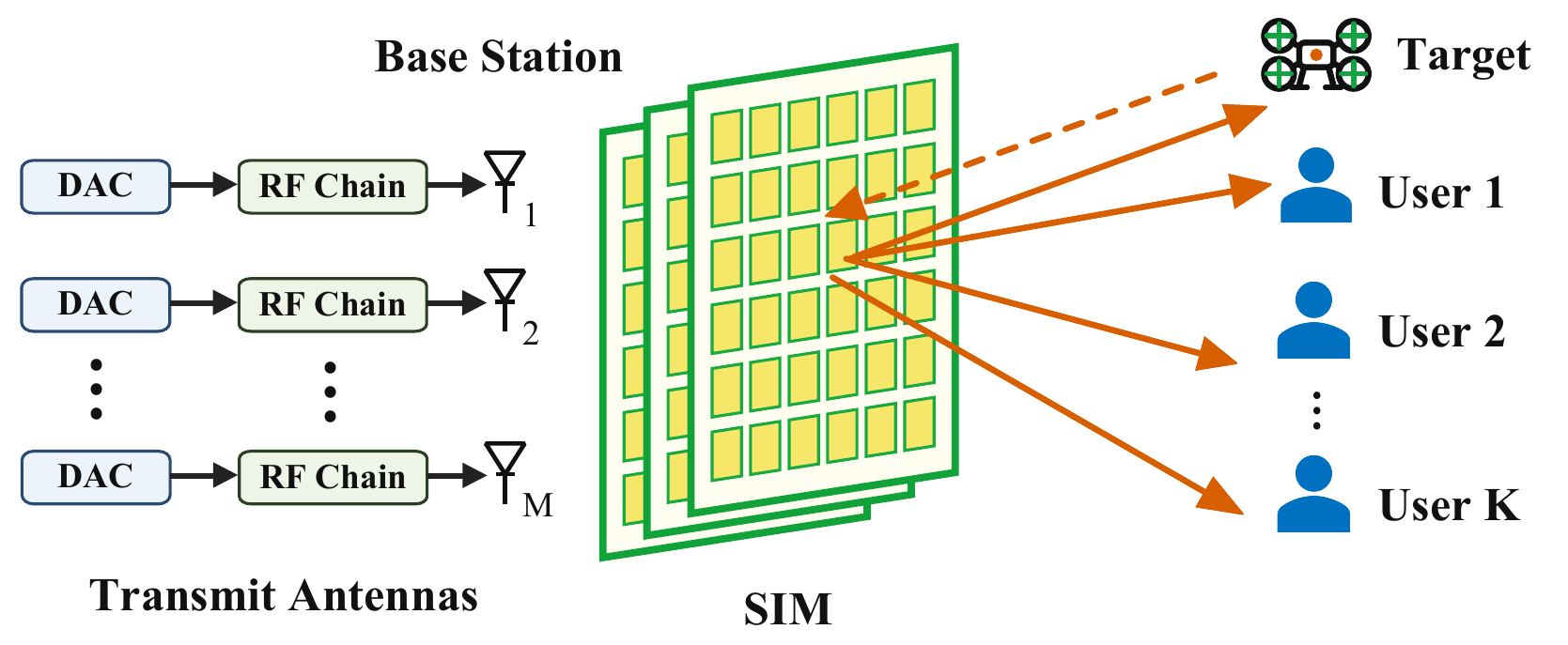}
	\caption{Models of the considered SIM-aided ISAC system.}
	\label{fig:model}
	\vspace{-20pt}
\end{figure}

As shown in Fig.~\ref{fig:model}, we consider a discrete SIM-aided ISAC system, where a base station (BS) equipped with $L$-layer SIM, serving $K$ single-antenna users while simultaneously sensing a point-like target. The SIM is equipped in the $y\text{-}z$ plane, where each layer consists of $N = N_1 \times N_2$ reconfigurable meta-atoms and $N$-element sensing array co-located with the output SIM layer with $N$ receive RF chains. Let $\mathcal{L}=\{1,\ldots,L\}$, $\mathcal{M}=\{1,\ldots,M\}$, $\mathcal{N}=\{1,\ldots,N\}$, and $\mathcal{K}=\{1,\ldots,K\}$ denote the index sets of the SIM layers, transmit antennas, meta-atoms, and users. Based on the Rayleigh-Sommerfeld diffraction theory\cite{goodman2005introduction}, for $l\in\mathcal{L}\setminus\{1\}$ and $n, n'\in\mathcal{N}$, we define  $\mathbf{W}^l \in \mathbb{C}^{N \times N}$ as the propagation matrix from the $(l-1)$-th layer to the $l$-th layer, and the $(n, n')$-th entry is given by
\vspace{-6pt}
\begin{equation}
	w_{n,n'}^l = \frac{d_y d_z \cos\varphi_{n,n'}^l}{d_{n,n'}^l} \left(\frac{1}{2\pi d_{n,n'}^l} - \frac{j}{\lambda}\right) e^{j2\pi d_{n,n'}^l/\lambda},\label{eq:sim_initial}
	\vspace{-5pt}
\end{equation}
\noindent where $\lambda$ is the wavelength, $d_{n,n'}^l$ is the transmission distance, $\varphi_{n,n'}^l$ is the incident angle, $d_y$ and $d_z$ are the length of meta-atoms. Similarly, let $\mathbf{w}_{t}^{m}\in\mathbb{C}^{N\times 1}$ denote the propagation vector from the $m$-th transmit antenna to the first SIM layer. For the $l$-th layer, the phase shift matrix is denoted as 
$
	\bm{\Phi}^l = \text{diag}(\bm{\phi}^l) = \text{diag}\left(e^{j\phi_1^l}, e^{j\phi_2^l}, \ldots, e^{j\phi_N^l}\right) \in \mathbb{C}^{N \times N},
$
where $\phi_n^l \in
\left\{
0,\frac{2\pi}{2^B},\ldots,\frac{2\pi(2^B-1)}{2^B}
\right\}$ denotes the $B$-bit phase shift of the $n$-th meta-atom in the $l$-th layer. Therefore, the wave-domain beamforming matrix is then given by\cite{an2023sim_hmimo}
\vspace{-6pt}
\begin{equation}
	\mathbf{G}
	=
	\boldsymbol{\Phi}^{L}\mathbf{W}^{L}
	\boldsymbol{\Phi}^{L-1}\mathbf{W}^{L-1}
	\cdots
	\boldsymbol{\Phi}^{2}\mathbf{W}^{2}
	\boldsymbol{\Phi}^{1}
	\in\mathbb{C}^{N\times N}.
	\label{eq:sim_beamforming}
	\vspace{-5pt}
\end{equation}
\vspace{-14pt}

The BS transmits dual-functional signals for simultaneous communication and sensing. Assuming $M=K$, the signal incident on the first SIM layer at time $t$ is
$
\bm{x}[t]\!=\!\mathbf{W}_{t}\mathbf{P}\bm{s}[t]\in\mathbb{C}^{N\times 1},
$
where $\mathbf{W}_{t}\!=\![\mathbf{w}_{t}^1,\ldots,\mathbf{w}_{t}^M]\!\in\!\mathbb{C}^{N\times M}$ is the transmit-to-SIM propagation matrix, and $\bm{s}[t]\!\in\!\mathbb{C}^{K\times 1}$ is the symbol vector satisfying $\mathbb{E}\{\bm{s}[t]\bm{s}^{\rm H}[t]\}\!=\!\mathbf{I}_K$. The power allocation matrix is $\mathbf{P}\!=\!\operatorname{diag}(p_1,\ldots,p_K)\!\in\!\mathbb{R}^{K\times K}$, where $p_k$ denotes the transmit amplitude of the $k$-th stream and satisfies $\sum_{k\!=\!1}^{K}p_k^2\leq P_{\max}$, with $P_{\max}$ denoting the power budget. For a block of $T$ symbols,
$
\mathbf{X}\!=\!\mathbf{W}_{t}\mathbf{P}\mathbf{S}\!\in\!\mathbb{C}^{N\times T},
$
where $\mathbf{S}\!=\![\bm{s}[1],\ldots,\bm{s}[T]]\!\in\!\mathbb{C}^{K\times T}$. For sufficiently large $T$, $\frac{1}{T}\mathbf{S}\mathbf{S}^{\rm H}\approx\mathbf{I}_K$, and hence
$
\mathbf{R}_{\mathbf{X}}
\!\triangleq\!
\frac{1}{T}\mathbf{X}\mathbf{X}^{\rm H}
\!\approx\!
\mathbf{W}_{t}\mathbf{P}\mathbf{P}^{\rm H}(\mathbf{W}_{t})^{\rm H}
$\cite{liu2022crb_joint_radar_comm}.

\vspace{-12pt}
\subsection{Metrics for Communication}
\vspace{-3pt}
We consider the spatial correlated Rayleigh fading model $\bm{h}_k \sim \mathcal{CN}(0, \beta_k \mathbf{R}) \in \mathbb{C}^{N \times 1}$ from the last SIM layer to user $k$, where $\beta_k$ is the large-scale path loss and $\mathbf{R}$ is the channel correlation matrix. According to \cite{goldsmith2005wireless}, the $(n, n')$-th entry is
\vspace{-6pt}
\begin{equation}
	[\mathbf{R}]_{n,n'}
	=
	\operatorname{sinc}\left(
	\frac{2\left\|\bm{u}_{n}-\bm{u}_{n'}\right\|_{2}}{\lambda}
	\right),
	\quad n,n'\in\mathcal{N},
	\label{eq:spatial_correlation}
	\vspace{-6pt}
\end{equation}
where $\bm{u}_{n} = \left[ 0,\,
	\frac{\operatorname{mod}(n-1,N_{1})\lambda}{2},\,
	\frac{\left\lfloor (n-1)/N_{1}\right\rfloor\lambda}{2}
	\right]^{\mathrm{T}}$ denotes the position of the $n$-th meta-atom.

The signal received at user $k$ is denoted as
\vspace{-8pt}
\begin{equation}\label{eq:user_received_signal}
	\begin{aligned}
		y_k[t] \!=\! {\bm{h}_k^{\rm H} \mathbf{G} \mathbf{w}_{t}^{k} p_k s_k[t]} \!+\! {\sum_{i=1, i \neq k}^K \bm{h}_k^{\rm H} \mathbf{G}\mathbf{w}_{t}^{i} p_i s_i[t]}+ {n_k[t]},
	\end{aligned}
	\vspace{-5pt}
\end{equation}
\vspace{-3pt}

\noindent where $n_k[t] \sim \mathcal{CN}(0, \sigma_k^2)$ is the additive white Gaussian noise.

\vspace{-12pt}
\subsection{Metrics for Sensing}
\vspace{-5pt}
We model target direction $\bm{\theta} = [\theta_t, \vartheta_t]^{\rm T}$ with steering vector
$
		\bm{a}(\bm{\theta})
		= 
		\left[
		1, e^{-j\pi\sin\theta_t\sin\vartheta_t},
		\ldots,
		e^{-j\pi(N_1-1)\sin\theta_t\sin\vartheta_t}
		\right]^{\rm T} \otimes
		\left[
		1, e^{-j\pi\cos\theta_t},
		\ldots,
		e^{-j\pi(N_2-1)\cos\theta_t}
		\right]^{\rm T}\big/{\sqrt{N}},
$
where $\otimes$ denotes the Kronecker product, $\theta_t$ denotes elevation angle and $\vartheta_t$ denotes azimuth angle.
The radar echo signal is denoted as
\vspace{-5pt} 
\begin{equation}\label{eq:radar_received_signal}
	\mathbf{Y}_r = \alpha_r \bm{a}(\bm{\theta}) \bm{a}(\bm{\theta})^{\rm H} \mathbf{G} \mathbf{X} + \mathbf{N}_r \in \mathbb{C}^{N \times T},
	\vspace{-5pt}
\end{equation}

\noindent where $\alpha_r\in\mathbb{C}$ is the complex target reflection coefficient, including the radar cross section (RCS) and round-trip path loss, and 
$[\mathbf{N}_r]_{n,t}\sim\mathcal{CN}(0,\sigma_r^2)$ denotes the independent AWGN.
To derive the CRB for estimating $\bm\theta$, we first vectorize the received signal  $\bY_r$ as
\vspace{-5pt}
\begin{equation}\label{eq:vectorized_signal}
	\bm{y}_r = \text{vec}(\mathbf{Y}_r) = \alpha_r \text{vec}\left(\bm{a}(\bm{\theta}) \bm{a}(\bm{\theta})^{\rm H} \mathbf{G} \mathbf{W}_{t} \mathbf{P} \mathbf{S}\right) + \bm{n}_r,
	\vspace{-5pt}
\end{equation}
where $\bm{n}_r = \text{vec}(\mathbf{N}_r) \sim \mathcal{CN}(0, \sigma_r^2 \mathbf{I}_{NT})$.
In this case, the unknown parameter vector to be estimated is denoted as 
$\bm{\xi} = \left[ \btheta, \bm\alpha \right]^{\rm T} = [\theta_t, \vartheta_t, \text{Re}\{\alpha_r\}, \text{Im}\{\alpha_r\}]^{\rm T}$.

Applying the Slepian–Bangs formula\cite{liu2022crb_joint_radar_comm} to $\by_r \sim {\cal CN}\left( \alpha_r {\rm vec}\left(  \ba(\btheta) \ba(\btheta)^{\rm H} \bG \bW_{t} \bP \bS \right), \sigma_r^2 \bI_{NT} \right)  $, the $(i,j)$-th element of the Fisher information matrix (FIM) $\bF_{\rm IM} \in \mathbb{R}^{4\times 4} $ can be obtained as
\vspace{-5pt}
\begin{equation} \label{Fisher_matrix}
	\begin{split}
		\bF_{\rm IM}(i,j) = \frac{2}{\sigma_r^2} {\rm Re}\left\{ \frac{\partial \alpha_r {\rm vec}\left( \bf{B} \right)^{\rm H} }{\partial \xi_i}  \frac{\partial \alpha_r {\rm vec}\left(  \bf{B} \right)}{\partial \xi_j} \right\},
	\end{split}
	\vspace{-6pt}
\end{equation}
where $\mathbf{B} = \bm{a}(\bm{\theta}) \bm{a}(\bm{\theta})^{\rm H} \mathbf{G} \mathbf{W}_{t} \mathbf{P} \mathbf{S} \in \mathbb{C}^{N \times T}$.

To characterize the sensing accuracy in terms of
the angular parameters, we partition $\bF_{\rm IM}$ and $\bC$ into
$2\times2$ blocks as
\vspace{-5pt}
\begin{equation*}
	\bC = \left[
	\begin{array}{cc}
		\bC_{\btheta\btheta^{\rm T}} & \bC_{\btheta\bm\alpha^{\rm T}} \\
		\bC_{\bm\alpha\btheta^{\rm T}} & \bC_{\bm\alpha\bm\alpha^{\rm T}}
	\end{array}
	\right] = \left[
	\begin{array}{cc}
		\bF_{\btheta\btheta^{\rm T}} & \bF_{\btheta\bm\alpha^{\rm T}} \\
		\bF_{\btheta\bm\alpha^{\rm T}}^{\rm T} & \bF_{\bm\alpha\bm\alpha^{\rm T}}
	\end{array}
	\right]^{-1} = \bF_{\rm IM}^{-1},
	\vspace{-4pt}
\end{equation*}

Based on the A-optimality criterion~\cite{li2008range}, we consider minimizing trace of the CRB matrix for estimating $\btheta$, which is given by
\vspace{-5pt}
\begin{equation}
	\begin{split}
		{\rm CRB}_{\bm\theta} = \!{\rm Tr}\left\{ \!\left( \bF_{\btheta\btheta^{\rm T}} \!-\! \bF_{\btheta\bm\alpha^{\rm T}} \bF_{\bm\alpha\bm\alpha^{\rm T}}^{-1} \bF_{\btheta\bm\alpha^{\rm T}}^{\rm T} \right)\!^{\!-1} \right\}.
	\end{split}
\end{equation}
\vspace{-2pt}
Although the above FIM focuses on angular parameters, Doppler can be incorporated as an additional parameter for moving targets while retaining the two-stage framework.
\vspace{-10pt}

\section{Two-Stage Precoding Design}
\vspace{-3pt}
\label{s3}

Directly optimizing the phase shifts and power allocation for CRB minimization is difficult due to the resulting highly non-convex formulation, discrete phase-shift constraints, and QoS constraints. To address this difficulty, we develop a two-stage optimization framework. Specifically, we first optimize an ideal precoder 
$\mathbf Q = \left[
\bm q_{1},
\bm q_{2},
\ldots,
\bm q_{K}
\right] \!\! \in \! \mathbb C^{N\times K}$
to obtain a sensing-oriented precoding reference. Then, the practical precoder 
$\mathbf{G}\mathbf{W}_t\mathbf{P}$ 
is optimized to approximate $\mathbf{Q}$ under the discrete phase-shift, transmit-power, and QoS constraints.

\vspace{-12pt}
\subsection{Ideal Precoder Design}
\vspace{-3pt}
After eliminating the nuisance reflection coefficient, the ideal precoder problem is formulated as
\vspace{-5pt}
\begin{equation} 
	\begin{split}
		\min_{ \bQ } ~ &{\rm Tr}\left\{ \left( \bF_{\btheta\btheta^{\rm T}} - \bF_{\btheta\bm\alpha^{\rm T}} \bF_{\bm\alpha\bm\alpha^{\rm T}}^{-1} \bF_{\btheta\bm\alpha^{\rm T}}^{\rm T} \right)^{-1} \right\} \\
		\mbox{s.t.}&
		~ \left\| \bQ \right\|_{\rm F}^2 \leq P_{\rm max}.
		\vspace{-5pt}
	\end{split}
	\label{CRB_prob}
\end{equation}

Following the standard epigraph reformulation\cite{liu2022crb_joint_radar_comm}, we introduce the auxiliary matrix $\bU \in \mathbb{R}^{2\times 2}$ satisfying $\bF_{\btheta\btheta^{\rm T}} - \bF_{\btheta\bm\alpha^{\rm T}} \bF_{\bm\alpha\bm\alpha^{\rm T}}^{-1} \bF_{\btheta\bm\alpha^{\rm T}}^{\rm T} \succeq \bU$ to transform the problem \eqref{CRB_prob} into
\vspace{-5pt}
\begin{equation} \label{CRB_prob_1}
	\begin{split}
		\min_{ \bQ, \bU } &~{\rm Tr}\left\{ \bU^{-1} \right\}    \\
		\mbox{s.t.}
		&~ \bU \!\succ\! 0, \| \bQ \|_F^2 \!\leq\! P_{\rm max},  \left[
		\begin{array}{cc}
			\!\!\!\bF_{\btheta\btheta^{\rm T}} \!\!-\!\! \bU &\!\! \bF_{\btheta\bm\alpha^{\rm T}}\!\!\! \\
			\!\!\!\bF_{\btheta\bm\alpha^{\rm T}}^{\rm T} &\!\! \bF_{\bm\alpha\bm\alpha^{\rm T}}\!\!\!
		\end{array}
		\right] \!\!\succeq\! 0.
		\vspace{-6pt}
	\end{split}
\end{equation}

In order to solve problem~\eqref{CRB_prob_1}, we propose a tractable optimization algorithm by introducing an auxiliary matrix
$\mathbf J\triangleq\sum_{k=1}^K\mathbf J_k$, where $\mathbf J_k\triangleq\bm q_k\bm q_k^{\rm H}\succeq\mathbf 0\in\mathbb C^{N\times N}$. 
Therefore, the ideal precoder design is reformulated as
\vspace{-5pt}
\begin{subequations}
	\label{prob:ideal_covariance}
	\begin{align}
		\min_{\{\mathbf J_k\}_{k\in\mathcal K},\,\mathbf U}\quad
		&\operatorname{Tr}\{\mathbf U^{-1}\}\\
		{\rm s.t.}\quad
		&\mathbf J_k\succeq\mathbf 0,\,k\in\mathcal K,\ \mathbf U\succ\mathbf 0,\ \operatorname{Tr}(\mathbf J)\leq P_{\max},\\
		&\begin{bmatrix}
			\mathbf F_{\boldsymbol\theta\boldsymbol\theta^{\rm T}}(\mathbf J)-\mathbf U & \mathbf F_{\boldsymbol\theta\boldsymbol\alpha^{\rm T}}(\mathbf J)\\
			\mathbf F_{\boldsymbol\theta\boldsymbol\alpha^{\rm T}}^{\rm T}(\mathbf J) & \mathbf F_{\boldsymbol\alpha\boldsymbol\alpha^{\rm T}}(\mathbf J)
		\end{bmatrix}\succeq\mathbf 0.
	\end{align}
	\vspace{-8pt}
\end{subequations}




Problem~\eqref{prob:ideal_covariance} is a convex SDP and can be solved by CVX~\cite{cvx}. 
This problem has a similar structure to that in~\cite{liu2022crb_joint_radar_comm}, and the SDR is tight in terms of the optimal objective value. 
Therefore, a rank-one optimal solution $\{\mathbf J_k^\star\}_{k\in\mathcal K}$ exists, and $\bm q_k$ can be recovered from $\mathbf J_k^\star$ via rank-one decomposition.

\vspace{-12pt}
\subsection{Practical SIM Precoder Realization}
\vspace{-5pt}

In the second stage, we optimize the multi-layer SIM phase shifts and power allocation to minimize the precoding mismatch between the practical SIM precoder $\mathbf G\mathbf W_t\mathbf P$ and the ideal precoder $\mathbf Q$, thereby approaching the sensing performance of the ideal design under practical SIM constraints.
\vspace{-12pt}
	\begin{align}
		\min_{ \{ \bm\Phi^l \}_{l=1}^L, \bP } &~\left\| \bG \bW_{t} \bP - \bQ \right\|_{\rm F}^2  \notag  \\
		\mbox{s.t.}\quad
		&~ {\cal C}_{\Phi}: \bPhi^l \!=\! \diag(\bm\phi^l),~\phi_n^l \!=\! e^{j \phi_n^l},~l \!\in \!{\cal L},\,n \!\in\! {\cal N},  \notag\\
		&~ {\cal C}_{\phi}: \phi_n^l \in
		\left\{
		0,\frac{2\pi}{2^B},\ldots,\frac{2\pi(2^B-1)}{2^B}
		\right\}\\
		&~ {\cal C}_{p}: \sum_{k=1}^{K} p_k^2 \leq P_{\rm max}, ~ {\cal C}_{0}: p_k \geq 0,\, k \in {\cal K}, \notag\\
		&~ {\cal C}_{k}: \frac{ | \bh_k^{\rm H} \bG \mathbf{w}_t^k p_k  |^2}{\sum_{i=1,i\neq k}^{K}\!| \bh_k^{\rm H} \bG \mathbf{w}_t^i p_i  |^2\! + \!\sigma_k^2   } \!\geq\! \gamma_{\rm th}.\notag
	\end{align}
	\vspace{-5pt}

\subsubsection{Power Allocation Optimization}

For given SIM phase shifts $\{\bm\Phi^l\}_{l=1}^{L}$, the power allocation problem is formulated as
\vspace{-8pt}
\begin{equation}
	\label{prob:power_original}
		\underset{\bP}{\operatorname{min}} \ 
		\sum_{k=1}^{K}
		\left\|
		\bG \mathbf{w}_t^k p_k - \bm q_k
		\right\|_2^2
		\quad
		\mbox{s.t.} \ 
		{\cal C}_{p}, {\cal C}_{0}, {\cal C}_{k}.
		\vspace{-5pt}
\end{equation}

After the preceding reformulation, the objective function and the power constraint 
become convex. However, the constraints ${\cal C}_{k}$ are non-convex. For notational simplicity, define the effective channel gain from the $i$-th stream to the $k$-th user as
\vspace{-7pt}
\begin{equation}
	\eta_{k,i}
	\triangleq
	\left|
	\bh_k^{\rm H}\bG\mathbf{w}_t^i
	\right|^2
	=
	\operatorname{Re}
	\left\{
	\mathbf{w}_t^{i,\rm H}
	\bG^{\rm H}
	\bh_k\bh_k^{\rm H}
	\bG
	\mathbf{w}_t^i
	\right\}.
	\label{eq:eta_definition}
	\vspace{-5pt}
\end{equation}

To obtain a tractable convex approximation, we first propose a first-order upper approximation 
\vspace{-5pt}
\begin{equation}
	\gamma_{\rm th}
	\sum_{\substack{i=1, i\neq k}}^{K}
	\eta_{k,i}p_i^2
	-
	\eta_{k,k}
	\left(
	2\hat p_kp_k-\hat p_k^2
	\right)
	+
	\gamma_{\rm th}\sigma_k^2
	\leq 0.
	\label{eq:sinr_power_sca}
	\vspace{-5pt}
\end{equation}

The resulting QCQP problem can be solved by CVX.

\subsubsection{Phase-Shift Optimization}

Given $\bP$ and the other SIM phase shifts, we update $\bm\phi^l$ of the $l$-th layer and rewrite the transmit matrix as
$
\bG\mathbf{w}_t^kp_k = \bC_k^l\bm\phi^l,
$
where $\bC_k^l$ is the equivalent coefficient matrix of the $l$-th SIM layer, given by
\vspace{-5pt}
\begin{align}
	\bC_k^1
	&=
	\bPhi^L\bW^L\cdots\bPhi^2\bW^2
	\diag\!\left(\mathbf{w}_t^k p_k\right),
	\notag \\
	\bC_k^L
	&=
	\diag\!\left(
	\bW^L\bPhi^{L-1}\bW^{L-1}\cdots\bPhi^2\bW^2\bPhi^1
	\mathbf{w}_t^k p_k
	\right),
	\label{eq:C_i_l_def} \\
	\bC_k^l
	&=
	\bPhi^L\bW^L\cdots\bPhi^{l+1}\bW^{l+1}
	\diag\!\left(
	\bW^l\bPhi^{l-1}\cdots\bPhi^1
	\mathbf{w}_t^k p_k
	\right),
	\notag
		\vspace{-10pt}
	\end{align}
	where $l=2,\ldots,L-1.$

Therefore, the phase shift optimization subproblem for the $l$-th SIM layer can be formulated as
\vspace{-5pt}
\begin{equation}
	\label{prob:phi_original}
		\underset{\bm\phi^l}{\operatorname{min}} \quad
		 \sum_{k=1}^{K}
		\left\|
		\bC_k^l\bm\phi^l-\bm q_k
		\right\|_2^2,
		\quad
		\mbox{s.t.} \ 
		{\cal C}_{\Phi}, {\cal C}_{\phi}, {\cal C}_{k}.
		\vspace{-5pt}
\end{equation}

Problem~\eqref{prob:phi_original} is challenging due to the combinatorial discrete phase-shift constraint and non-convex QoS constraints. We thus introduce an auxiliary vector $\boldsymbol{\psi}^{l} \in \mathbb{C}^{N^2}$ and solve the problem through the following iterative updates
\vspace{-6pt}
\begin{subequations}\label{eq:admm_updates}
	\begin{align}
		\boldsymbol{\psi}^{l,t+1} &= 
		\underset{\boldsymbol{\psi}^{l}}{\operatorname{argmin}} \;\; 
		\mathcal{J}(\boldsymbol{\psi}^{l}, 
		\boldsymbol{\Phi}^{l,t}, \boldsymbol{\kappa}^t),
		\label{eq:admm_theta} \\
		\boldsymbol{\Phi}^{l,t+1} &= 
		\underset{\boldsymbol{\Phi}^l \in\, 
			\mathcal{C}_{\boldsymbol{\Phi}} \cap \mathcal{C}_{\boldsymbol{\phi}}}{\operatorname{argmin}} \;\;
		\|\mathrm{vec}(\boldsymbol{\Phi}^{l}) - \boldsymbol{\psi}^{l,t+1} - \boldsymbol{\kappa}^t\|_2^2,
		\label{eq:admm_phi} \\
		\boldsymbol{\kappa}^{t+1} &= \boldsymbol{\kappa}^t 
		+ \boldsymbol{\psi}^{l,t+1} 
		- \mathrm{vec}(\boldsymbol{\Phi}^{l,t+1}).
		\label{eq:admm_dual}
		\vspace{-5pt}
	\end{align}
\end{subequations}
where $\mathcal{J} =\sum_{k=1}^{K}
		\left\|
		\bC_k^l\bm\phi^l-\bm q_k
		\right\|_2^2 + \frac{\mu}{2}\|\boldsymbol{\psi}^{l} - \mathrm{vec}(\boldsymbol{\Phi}^{l}) + \boldsymbol{\kappa}\|_2^2$,
$\boldsymbol{\kappa}\!\in\!\mathbb{C}^{N^2}$ is the scaled dual variable, and
following the standard ADMM scheme, the penalty parameter $\mu$ kept constant.

For the $\boldsymbol{\psi}$-subproblem~\eqref{eq:admm_theta}, we approximate constraint ${\cal C}_{k}$ by the following first-order lower bound
\vspace{-5pt}
\begin{equation}
	\begin{aligned}
		&\gamma_{\rm th}
		\sum_{\substack{i=1,\, i\neq k}}^{K}
		(\bm\phi^l)^{\rm H}
		\bD_{k,i}^l
		\bm\phi^l
		-
		2\operatorname{Re}
		\left\{
		(\hat{\bm\phi}^l)^{\rm H}
		\bD_{k,k}^l
		\bm\phi^l
		\right\}  \\
		&\quad
		+
		(\hat{\bm\phi}^l)^{\rm H}
		\bD_{k,k}^l
		\hat{\bm\phi}^l
		+
		\gamma_{\rm th}\sigma_k^2
		\leq 0,
		\quad k\in{\cal K}.
	\end{aligned}
	\label{eq:phi_sca_lower_bound}
	\vspace{-5pt}
\end{equation}
where $
\bD_{k,i}^l \!\!
\triangleq \!\!
(\bC_i^l)^{\rm H}
\bh_k\bh_k^{\rm H}
\bC_i^l.
\label{eq:D_definition}
$ 
Then, subproblem~\eqref{eq:admm_theta} is convex and can be solved by CVX. 

For $\boldsymbol{\Phi}$-subproblem~\eqref{eq:admm_phi}, we propose
the following projection rule for discrete phase shift update.

\textbf{Proposition 1.} For a given
$
z_n^l=\left[(\bm\psi^l)^{t+1}-\bm\kappa^l\right]_n
$,
the optimal discrete projection of $z_n^l$ is given by
\vspace{-5pt}
\begin{equation}
	\left(\phi_n^l\right)^{t+1}
	=
	\begin{cases}
		e^{j\frac{(\tau-1)\pi}{2^{B-1}}},
		& \text{if }~
		\angle\!\left(z_n^l\right)
		<
		\frac{(2\tau-1)\pi}{2^B},
		\\[1mm]
		e^{j\frac{\tau\pi}{2^{B-1}}},
		& \text{if }~
		\angle\!\left(z_n^l\right)
		\geq
		\frac{(2\tau-1)\pi}{2^B},
	\end{cases}
	\label{eq:discrete_projection}
	\vspace{-5pt}
\end{equation}
where
$
	\tau
	=
	\left\lfloor
	\frac{
		2^{B-1}\angle\!\left(z_n^l\right)+\pi
	}{\pi}
	\right\rfloor
$, and \(\lfloor \cdot \rfloor\) denotes floor operation.

\textit{Proof:}
Since $\left|\phi_n^l\right|=1$, minimizing
$\left|\phi_n^l-z_n^l\right|^2$ is equivalent to maximizing
$\Re\{(\phi_n^l)^*z_n^l\}$. Therefore, the optimal solution is obtained by
selecting the discrete phase closest to $\angle(z_n^l)$. The threshold
$\frac{(2\tau-1)\pi}{2^B}$ is the midpoint between two adjacent discrete
phase levels, which completes the proof.

\vspace{-12pt}
\subsection{Computational Complexity}
\vspace{-3pt}
The first-stage SDR-based ideal precoder design has a computational
complexity of $\mathcal{O}\left(N^{6.5}\right)$.
In the second stage, the power-allocation subproblem has a computational
complexity of $\mathcal{O}\left(K^{3.5}\right)$, while each layer-wise
continuous phase-shift update has a computational complexity of
$\mathcal{O}\left(\sqrt{N+K}\left(N^3+KN^2\right)\right)$.
Moreover, the closed-form discrete phase projection has a computational
complexity of $\mathcal{O}(N)$.
Therefore, the computational complexity of the proposed
two-stage algorithm is
$\mathcal{O}\left(N^{6.5}+K^{3.5}+L\left[\sqrt{N+K}\left(N^3+KN^2\right)+N\right]\right).$

\vspace{-10pt}
\section{Simulation Results}
\vspace{-3pt}
In this section, numerical simulations are conducted to evaluate the proposed
CRB minimization design for the SIM-aided ISAC system. The carrier frequency
is $f_c \!=\!10$ GHz, and the spacing between adjacent SIM meta-atoms is
$d_z = d_y \!=\!\lambda/2$. The total SIM thickness is fixed at $5\lambda$, with inter-layer spacing $5\lambda/L$. We set $T\!=\!128$ sensing snapshots and sensing noise power $\sigma_r^2\!=\!-80$ dBm. Perfect instantaneous user CSI is assumed at the BS. Inter-layer propagation loss is captured by~\eqref{eq:sim_initial}, while additional meta-atom insertion loss is neglected.
The users are randomly located $50$-$80$ m from the BS, with large-scale path loss $\beta_k=\beta_0 d_k^{-2.2}$, where \(\beta_0\) denotes the path loss and $d_k$ denotes the distance between BS and user.

\begin{figure}[!t]
	\vspace{-15pt}
	\centering
	\subfloat[\scriptsize Convergence of the algorithm.]
	{
		\includegraphics[width=0.45\columnwidth]{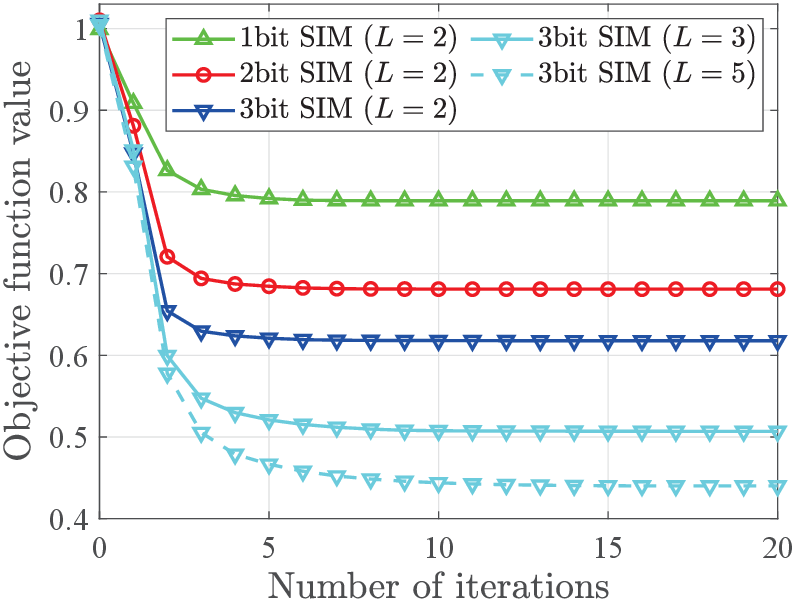}
		\label{fig:convergence}
	}
	\hfill
	\subfloat[\scriptsize ISAC performance trade-off.]
	{
		\includegraphics[width=0.45\columnwidth]{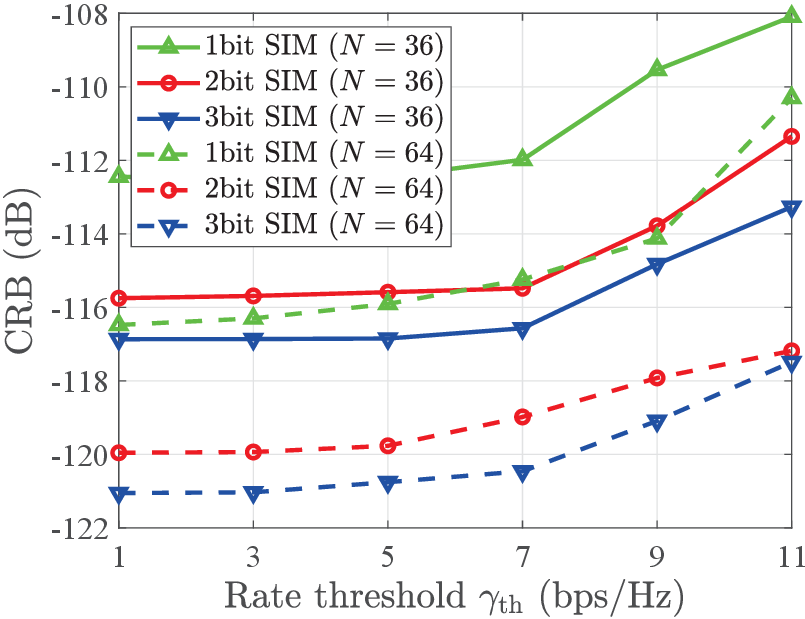}
		\label{fig:crb_tradeoff}
	}
	\caption{Convergence behavior of the proposed algorithm and ISAC performance trade-off.}
	\label{fig:convergence_crb_2}
	\vspace{-25pt}
\end{figure}

\begin{figure}[!t]
	\centering
	\subfloat[\scriptsize CRB versus number of layers.]
	{
		\includegraphics[width=0.45\columnwidth]{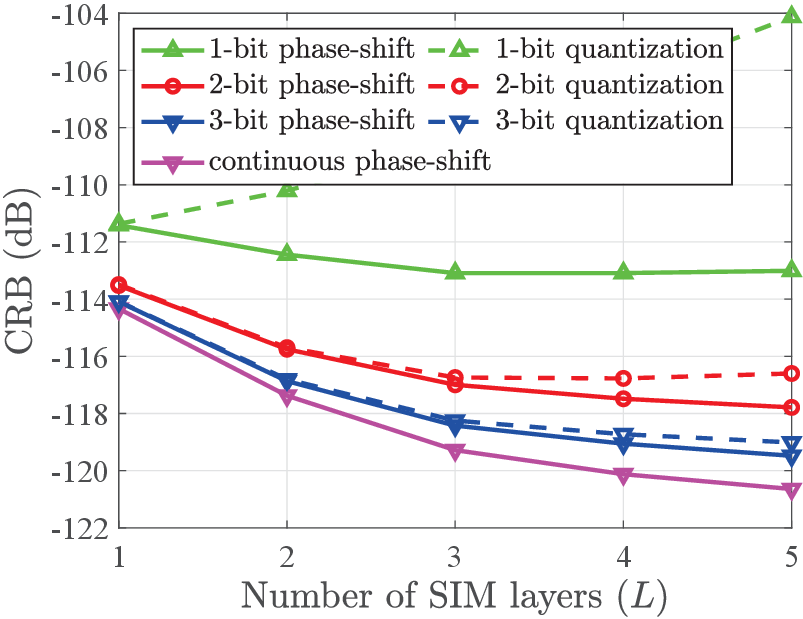}
		\label{fig:crb_L}
	}
	\hfill
	\subfloat[\scriptsize CRB versus transmit power.]
	{
		\includegraphics[width=0.45\columnwidth]{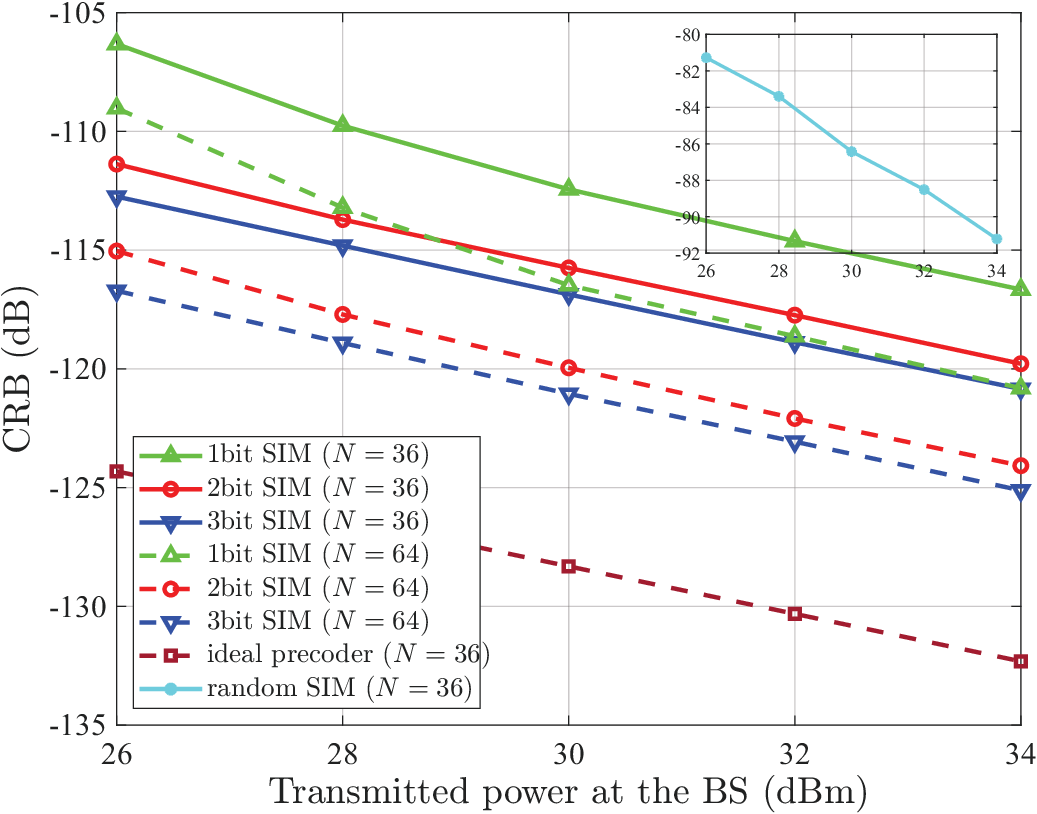}
		\label{fig:crb_p}
	}
	\caption{CRB performance comparison of the proposed algorithm versus SIM layers and transmit power.}
	\label{fig:crb_3}
	\vspace{-25pt}
\end{figure}

\begin{figure}[!t]
	\centering
	\subfloat[\scriptsize Elevation response for angle $\theta_t$.]
	{
		\includegraphics[width=0.45\columnwidth]{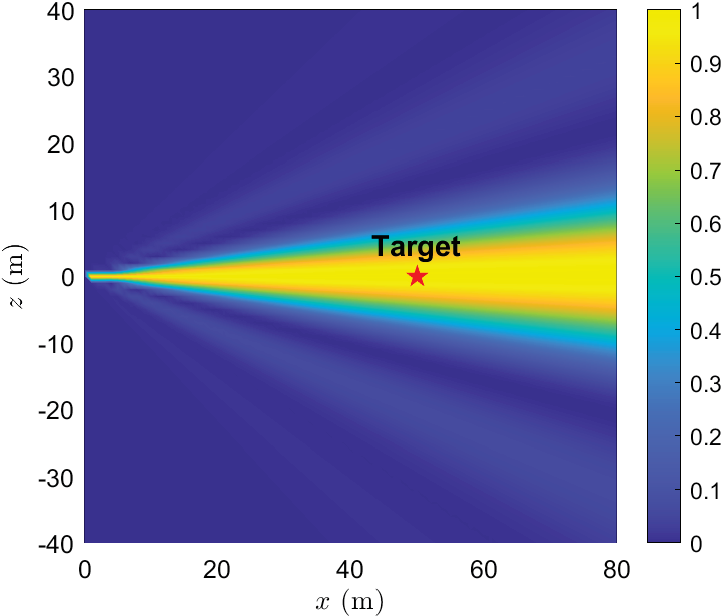}
		\label{fig:theta_result}
	}
	\hfill
	\subfloat[\scriptsize Azimuth response for angle $\vartheta_t$.]
	{
		\includegraphics[width=0.45\columnwidth]{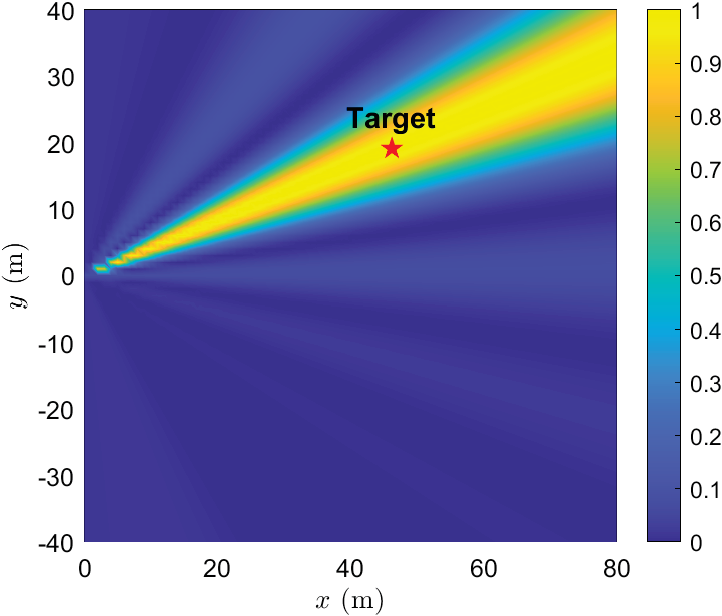}
		\label{fig:phi_result}
	}
	\caption{Normalized sensing response of the proposed SIM-aided ISAC scheme.}
	\label{fig:angle_results}
	\vspace{-20pt}
\end{figure}

Fig.~\ref{fig:convergence_crb_2} evaluates the convergence behavior and the ISAC performance trade-off.
As shown in Fig.~\ref{fig:convergence_crb_2}\subref{fig:convergence}, the objective value decreases rapidly in the first few iterations and then gradually converges, demonstrating the stable convergence of the proposed algorithm.
Fig.~\ref{fig:convergence_crb_2}\subref{fig:crb_tradeoff} shows that the CRB increases with the rate threshold, since stricter communication requirements allocate more spatial and power resources to communication, thereby degrading sensing accuracy.
Moreover, increasing the number of SIM meta-atoms and the phase-shift resolution consistently reduces the CRB, indicating that additional spatial degrees of freedom and finer phase control can alleviate the sensing loss caused by stringent QoS requirements.

Fig.~\ref{fig:crb_3} compares the CRB performance under different SIM configurations and transmit powers.
As shown in Fig.~\ref{fig:crb_3}\subref{fig:crb_L}, increasing the number of SIM layers reduces the CRB due to the enhanced wave-domain beamforming capability.
Moreover, the continuous phase-shift scheme achieves the lowest CRB, while the proposed discrete phase-shift design consistently outperforms direct phase quantization, demonstrating the benefit of explicitly optimizing the discrete phases rather than simply quantizing.
Fig.~\ref{fig:crb_3}\subref{fig:crb_p} shows that the CRB decreases with the BS transmit power and the number of meta-atoms. The ideal precoder provides a lower bound, whereas the proposed SIM design outperforms the random  scheme, showing that the gap mainly stems from practical SIM realization.

Fig.~\ref{fig:angle_results} illustrates the spatial beampatterns for angle estimation. It can be observed that the sensing response exhibits a dominant high-gain region around the target direction marked by the red star. This indicates that the proposed SIM-aided architecture is capable of concentrating the reflected sensing energy toward the desired angular direction. In addition, the responses outside the main direction remain relatively weak, which confirms the effectiveness of the optimized phase shifts in suppressing undesired angular components.

\vspace{-10pt}
\section{conclusion}
\vspace{-3pt}
In this paper, CRB minimization for a discrete SIM-aided multiuser ISAC system was investigated. To address the resulting non-convex problem, a two-stage transmit design framework was developed, where an ideal precoder is first obtained and then synthesized on the practical multi-layer SIM through alternating power allocation and phase-shift updates. Simulation results demonstrated fast convergence, revealed the sensing-communication trade-off, and verified the performance gains of increasing SIM layers and phase-shift resolution.
\vspace{-8pt}
\bibliographystyle{IEEEtran}
\bibliography{refs}

\end{document}